\documentclass[amssymb,prb,twocolumn,showpacs,reprint]{revtex4}
\usepackage{amsmath}
\usepackage{graphicx}
\usepackage{verbatim}
\usepackage{graphics}

\newcommand{\bea}{\begin{eqnarray}}
\newcommand{\eea}{\end{eqnarray}}
\newcommand{\be}{\begin{equation}}
\newcommand{\ee}{\end{equation}}

\begin{document}

\title{Optimal energy quanta to current conversion}
\author{Rafael S\'anchez}
\author{Markus B\"uttiker}
\affiliation{D\'epartement de Physique Th\'eorique,
Universit\'e de Gen\`eve, CH-1211 Gen\`eve 4, Switzerland}
\date{\today}

\begin{abstract}
We present a microscopic discussion of a nano-sized structure which uses the quantization of energy levels and the physics of single charge Coulomb interaction to achieve an optimal conversion of heat flow to directed current. In our structure the quantization of energy levels and the Coulomb blockade lead to the transfer of quantized packets of energy from a hot source into an electric conductor to which it is capacitively coupled.
The fluctuation generated transfer of a single energy quantum translates into the directed motion of a single electron. 
Thus in our structure the ratio of the charge current to the heat current is determined by the ratio of the charge quantum to the energy quantum. 
An important novel aspect of our approach is that the direction of energy flow and the direction of electron motion are decoupled. 
\end{abstract}
\maketitle

\section{Introduction}
Recently thermal and thermoelectric transport phenomena have found increasing attention in the scientific community.  A particularly interesting task is the harvesting of energy from fluctuating environments to gain power for devices which are not permanently coupled to power sources. Our interest is in small mesoscopic structures which are well controlled and can be used to investigate basic aspects of thermoelectric transport phenomena. In small scale systems 
fluctuations are always present and significant compared to the average behavior. Channeling environmental fluctuations in a controlled way allows for instance to generate an electric current by converting environmental energy into directed motion. 

When the components of circuits are reduced to the nanoscale, quantum physics becomes important. For instance, energy is discrete in quantum dots so transport spectroscopy shows narrow resonances. 
In the mesoscopic regime, a set of pioneering thermoelectric experiments came with the work of Molenkamp {\it et al.}~ \cite{molenkamp,molenkamp2,molenkamp3}.
There, the transport response to temperature gradients created through a quantum point contact \cite{molenkamp,molenkamp2} and quantum dot \cite{molenkamp3} is measured. 
Recently circuit elements that manipulate heat flows rather than electric currents have been proposed or demonstrated in systems of reduced dimensionality~\cite{giazotto}, including rectifiers \cite{chang, scheibner,ming}, pumps~\cite{moskalets,arrachea,bauer} or refrigerators\cite{edwards,smith} that can approach the quantum limit \cite{pekola,pekolaHekking}. 

\begin{figure}[b]
\includegraphics[width=\linewidth,clip]{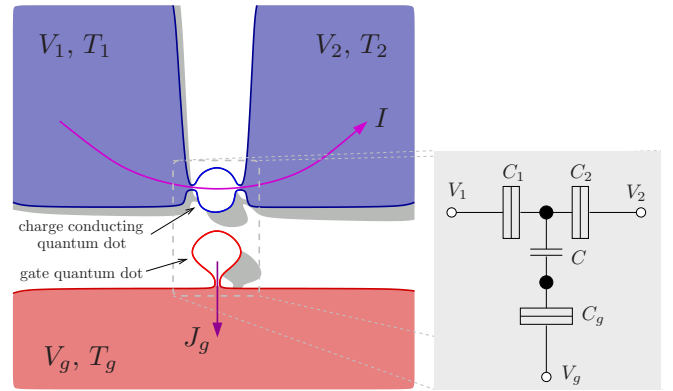}
\caption{\label{sys} Energy to current converter. 
The conductor, a quantum dot open to transport between two fermionic reservoirs at voltages $V_1$ and $V_2$ and temperatures $T_1$ and $T_2$, is coupled capacitively to a second dot which acts as a fluctuating gate coupled to a reservoir at voltage $V_g$ and temperature $T_g$. 
Here we discuss the case $T_1=T_2=T_s$.
}
\end{figure}

We consider a conceptually simple system which however turns out to be a laboratory for many (even counter intuitive) thermoelectric effects depending only on how different parameters are chosen.  Importantly among them, our device allows energy to work conversion at the highest efficiency. A quantum dot is coupled to two reservoirs  via two tunnel contacts which permit carrier exchange and is coupled capacitively to a gate such that there is only energy exchange between the conductor and the gate but remarkably no particle exchange.  The gate is itself structured into a quantum dot that permits carrier exchange with its reservoir. Thus there are two islands (dots) which interact only through the long range Coulomb force (see Fig. \ref{sys}).  To be specific, here we take the transmission through the tunnel barriers to be sufficiently small such that transport is defined by sequential tunneling of single electrons. Then, the dynamics of the system can be described by a master equation\cite{averin,cb}.
If intradot Coulomb repulsion is strong enough, the number of extra electrons in each quantum dot fluctuates between zero and one. The probability to find two extra electrons in one quantum dot is negligible. In such a configuration, the spin of the electron can be ignored. 

Quantum dots with the required properties~\cite{qdots} have been explored in metallic grains, semiconductor two dimensional electron gases and recently in nanowire heterostructures where the charging energy and the level spacing can be controlled~\cite{bjork}. These two energy scales constitute an upper bound to the temperature range for other thermoelectric quantum dot devices where heat is transported together with charge~\cite{edwards}. Our mechanism depends on the charge occupation of the quantum dots, so only charging energy is a relevant scale. Semiconductor quantum dots have typically charging energies which are an order of magnitude larger than the level spacing. Larger charging energies can be obtained in molecular structures. 

If the two dots are far from each other, they can be bridged to nevertheless obtain a strong coupling~\cite{bridge,hubel} at the same time ensuring good thermal isolation between the system and gate reservoirs. 
Effectively we have a three lead system with three independent reservoirs. The case of a four terminal structure in which each dot is coupled to two reservoirs has been the subject of a separate work by the two authors in collaboration with R. L\'opez and D. S\'anchez \cite{drag}.  Such a four terminal configuration permits in particular to investigate the effect of non-equilibrium noise due to current flow through one dot on the other dot with voltages maintained at equilibrium. Here we are concerned only with thermal equilibrium noise albeit with different temperatures at the reservoirs of the two dots.  For this reason it is sufficient to consider a dot connected to its reservoir only with one lead.  The dot connected to one lead then plays the role of a gate that is either hot or cold compared to the quantum dot with two leads, which we will call the conductor. 
The system is in an equilibrium state if there are no electrical or thermal current flows. 
In general, this requires that voltages $V_{1}$ and $V_{2}$ are equal and importantly requires that the temperatures of all three reservoirs are the same.
However, as we show below, special configurations allow to balance the presence of temperature gradients by applying finite voltages~\cite{humphrey}. 

The Coulomb coupling considered here is important also because it sets a limit on the close packing of electrical circuits. The denser circuits are packed the more important are the effects of charge fluctuations. Fluctuations of charge in one component can change such fundamental properties as detailed balance in another component of the circuit \cite{drag}. Exchange of particles is not required. It is sufficient that nearby systems exchange energy through interaction. Breaking of detailed balance can lead to directed motion as soon as a spatial symmetry is broken of either the system~\cite{vandenbroeck} or in the fluctuation generating component \cite{markus,blanter,vankampen,landauer,blmotor,blmotor2}.  We emphasize that a structured bath (a quantum dot with discrete energy levels) is not needed to drive current through an unbiased dot. Even a hot phonon bath can lead to current in an unbiased dot \cite{Imry}. However, the structured bath, the gate with a quantum dot quantizes the energy transfer. 

Our manuscript is organized as follows. In Section~\ref{sec:model} we describe the details of our system and introduce the theoretical tools needed for the analytical solution. The fluctuation generated current mechanism is presented in Section~\ref{sec:hotspots} and the efficiency of the heat to charge conversion, in Section~\ref{sec:efficiency}. A discussion is given in Section~\ref{sec:discussion}. Technical aspects and a detailed derivation of the main results are given in the Appendix.

\begin{figure}
\begin{center}
\includegraphics[width=0.6\linewidth]{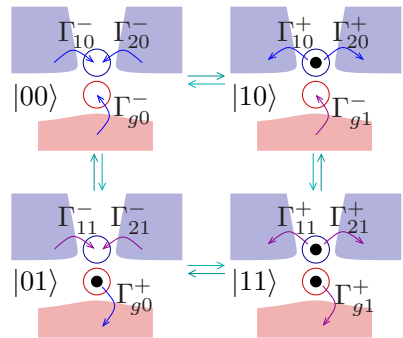}
\end{center}
\caption{\label{bas} \fontfamily{phv} Schematic representation of the four occupation states relevant for our system. The electron tunneling processes introducing or extracting an electron in the quantum dot system are described by the rates $\Gamma_{ln}^\pm$ through each barrier $l$, which depend on the charge occupation of the other quantum dot, $n=\{0,1\}$. Everytime a clockwise (anticlockwise) cycle, ${\cal C}_{+(-)}$,is completed, a quantum of energy $E_C$ is transferred from the gate into the conductor (and viceversa).}
\end{figure}
\section{Model}
\label{sec:model}
We are interested in non-equilibrium states and investigate the relation between the charge current flowing through the two terminals of the conductor, $I = I_2 = -I_1$, and the heat current $J_g$ flowing through the gate terminal at temperature $T_g$. The currents are defined as positive when flowing into the reservoirs.
The dynamical evolution of such a system is characterized by four states $|n_sn_g\rangle$, where $n_\alpha = \{0,1\}$ is the occupation number of each quantum dot, as sketched in Fig.~\ref{bas}. With the subindices $s$ and $g$ we denote the conductor and gate systems, respectively. The tunneling events are in general energy dependent. Due to the Coulomb interaction tunneling rates through terminal $l$ in one quantum dot, $\Gamma_{ln}$, are sensitive to the charge occupation, $n$, of the other quantum dot. Often the energy dependence of the tunneling rates is discussed only in terms of the energy dependence of the occupation functions (Fermi-Dirac functions).  However, here, in addition we require that either the density of states in the leads or the transmission through a tunnel junction depend on energy. This is natural since transmission probabilities depend typically in an exponential manner on the energy difference between the state out of which tunneling occurs and the barrier top. For our converter, the energy dependence of the transmission probabilities is absolutely essential. We will see that this energy dependence introduces the necessary asymmetry to get a directed current through the unbiased upper quantum dot. 

The capacitances associated with each tunnel junction (see Fig. \ref{sys}) define the charging energies, $U_{\alpha,n}(V_1,V_2,V_g)$, of each quantum dot, $\alpha$, depending on whether the other dot is empty ($n=0$) or occupied ($n=1$). They are calculated self-consistently in Appendix \ref{sec:elener}. When an electron tunnels into the empty system but leaves it only after a second electron has occupied the other quantum dot, 
a well defined energy  
\be
\label{ec}
E_C=U_{\alpha,1}-U_{\alpha,0}=\frac{2q^2}{\tilde C}
\ee
is exchanged between the two systems. Here, we have defined the total capacitance of each quantum dot $C_{\Sigma s}=C_1+C_2+C$ and $C_{\Sigma g}=C_g+C$, and the effective capacitance $\tilde C=(C_{\Sigma s}C_{\Sigma g}-C^2)/C$. The quantum of transferred energy $E_C$ depends only on the capacitance of the system and determines the heat flowing from one system to the other, as shown below. We emphasize that heat is transferred between the two systems due to electron-electron interaction. 

\subsection{Master equation}
\label{sec:meq}

We write a master equation for the density matrix, $\rho$, that represents the states of the quantum dot system. In the regime where transport is sequential, with $kT\gg\hbar\Gamma$ so broadening of the energy levels can be neglected, only the four diagonal terms describing the charge occupation probability of the system have to be taken into account~\cite{averin,cb}. 
In matricial form, the density matrix is a vector, $\rho=(\rho_{00},\rho_{10},\rho_{01},\rho_{11})$, so the master equation can be writen as $\dot\rho={\cal M}\rho$, with
\begin{eqnarray}\label{MLR}
\displaystyle
{
{\cal M}\!=\!\!
\left(\begin{array}{cccc}
-\Gamma_{\!s0}^--\Gamma_{\!g0}^- &\Gamma_{\!s0}^+& \Gamma_{\!g0}^+ & 0 \\
\Gamma_{\!s0}^- &  -\Gamma_{\!s0}^+-\Gamma_{\!g1}^- & 0 & \Gamma_{\!g1}^+ \\
\Gamma_{\!g0}^- & 0 & -\Gamma_{\!s1}^--\Gamma_{\!g0}^+ & \Gamma_{\!s1}^+ \\
0 & \Gamma_{\!g1}^- & \Gamma_{\!s1}^- & -\Gamma_{\!s1}^+-\Gamma_{\!g1}^+ \\
\end{array}  \right)
}
\end{eqnarray}
and $\Gamma_{\!sn}^\pm=\Gamma_{\!1n}^\pm+\Gamma_{\!2n}^\pm$. The rates $\Gamma_{ln}^\pm$ describe tunneling events that take an electron out ($+$) or into ($-$) the quantum dot system through junction $l$ when the other quantum dot contains $n=\{0,1\}$ electrons: $\Gamma_{\!ln}^-=\Gamma_{\!ln}f((E_{\alpha n}-qV_l)/kT_l)$, $\Gamma_{\!ln}^+=\Gamma_{\!ln}-\Gamma_{\!ln}^-$, with $E_{\alpha n}=\varepsilon_\alpha+U_{\alpha n}$, being $f(x)=(1+e^x)^{-1}$ the Fermi function. 
$\varepsilon_\alpha$ is the bare energy of the discrete level in quantum dot $\alpha$.

We are interested in the dc transport, which is given by the stationary solution of the master equation, ${\cal M}\bar\rho=0$. We can write it as:
\bea
\label{stat}
\bar\rho_{00}&=&\gamma^{-3}\sum_{\alpha=s,g}\sum_{i=\pm1}\sum_{n=0,1}\Gamma_{\alpha 0}^+\Gamma_{\alpha 1}^i\Gamma_{\bar ln}^+\delta_{|1-i|,2n}\nonumber\\
\bar\rho_{10}&=&\gamma^{-3}\sum_{i=\pm1}\sum_{n=0,1}\left(\Gamma_{s0}^-\Gamma_{s1}^i\Gamma_{gn}^++\Gamma_{g0}^i\Gamma_{g1}^+\Gamma_{sn}^-\right)\delta_{|1-i|,2n}\nonumber\\[-2.5mm]
&&\\[-2.5mm]
\bar\rho_{01}&=&\gamma^{-3}\sum_{i=\pm1}\sum_{n=0,1}\left(\Gamma_{s0}^i\Gamma_{s1}^+\Gamma_{gn}^-+\Gamma_{g0}^-\Gamma_{g1}^i\Gamma_{sn}^+\right)\delta_{|1-i|,2n}\nonumber\\
\bar\rho_{11}&=&\gamma^{-3}\sum_{\alpha=s,g}\sum_{i=\pm1}\sum_{n=0,1}\Gamma_{\alpha 0}^i\Gamma_{\alpha 1}^-\Gamma_{\bar \alpha n}^-\delta_{|1-i|,2n},\nonumber
\eea
with $\gamma^3=\sum_{\alpha in}\Gamma_{\alpha n}^i\left(\Gamma_{\alpha \bar n}^{\bar i}\Gamma_{\bar \alpha n}^{\bar i}+\Gamma_{\bar \alpha n}^{i}\sum_j\Gamma_{\bar \alpha n}^{j}\delta_{|1-i|,2n}\right)$ satisfying the normalization condition $\sum_i\bar \rho_{ii}=1$. The indices with a bar on top denote an opposite value, for example: $\bar s=g$, $\bar 0=1$.

The charge current through the conductor reads
\be
\label{eqi2}
I_l=q\sum_n\left(\Gamma_{ln}^+\bar\rho_{1n}-\Gamma_{ln}^-\bar\rho_{0n}\right),
\ee
while the heat currents are
\be
J_l=\sum_n(E_{sn}-qV_l)(\Gamma_{ln}^+\bar\rho_{1n}-\Gamma_{ln}^-\bar\rho_{0n}),
\ee
for terminals $l=1,2$ in the conductor, and 
\be\label{eqjq}
J_g=\sum_n(E_{gn}-qV_g)(\Gamma_{gn}^+\bar\rho_{n1}-\Gamma_{gn}^-\bar\rho_{n0}),
\ee
for the gate.
We can also write the energy currents as a combination of charge and heat currents: $W_l=J_l+V_lI_l$. Note that while charge and energy currents are conserved, so $\sum_lI_l=\sum_lW_l=0$, that is not the case for heat currents, due to the production of Joule heat in the presence of an external voltage. A finite heat is dissipated which we can write as
\be
\sum_lJ_l=\sum_l(V_i-V_l)I_l,
\ee
where, for simplicity, the voltage in terminal $i$ is considered as a reference.

\subsection{Quantum of transferred energy}
The relevant quantity is the heat exchanged between the two systems, $J_g$. From Eq. \eqref{stat} we can easily see that the terms in \eqref{eqjq} are related: 
\be
\Gamma_{g0}^+\bar\rho_{01}-\Gamma_{g0}^-\bar\rho_{00}=-\left(\Gamma_{g1}^+\bar\rho_{11}-\Gamma_{g1}^-\bar\rho_{10}\right).
\ee
Since $E_{g1}=E_{g0}+E_C$, we can write the heat current through the gate as:
\be
\label{jg}
J_g=-E_C\gamma^{-3}\left(\Gamma_{g0}^-\Gamma_{s1}^-\Gamma_{s0}^+\Gamma_{g1}^+-\Gamma_{s0}^-\Gamma_{g1}^-\Gamma_{s1}^+\Gamma_{g0}^+\right).
\ee
Note that only the terms of a {\it collision form} survive in the expression for the heat exchange between the systems and, more importantly, that it is proportional to $E_C$, meaning that the energy transferred between the two systems is quantized. In every cycle ${\cal C}_\pm$ (as defined in Fig.~\ref{bas}), 
an energy $\pm E_C$ is transferred from the gate into the conductor. The two terms on the right hand side of Eq. \eqref{jg} are proportional to the probablity of having one or the other of these cycles.

Interestingly, the quantum of transferred energy, $E_C$, can be estimated from transport spectroscopy measurements, as we show in Appendix~\ref{sec:ec}.

\section{Current from hot spots}
\label{sec:hotspots}
\begin{figure}
\includegraphics[width=\linewidth,clip]{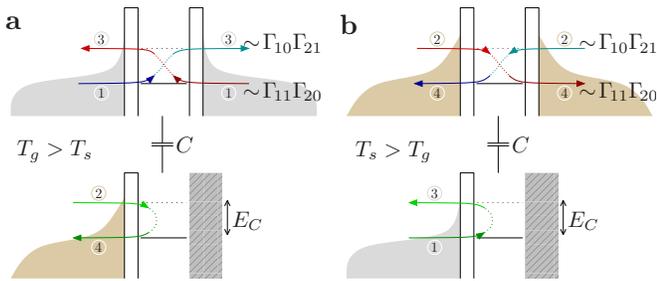}
\caption{\label{proc} 
(a) Energy diagram of the energy converter (when $V_1=V_2$ and $T_g > T_s$) showing the tunneling sequences that contribute to charge transport in the unbiased conductor. (b) If the temperature gradient is reversed ($T_g<T_s$), the fluctuation generated current will flow in the opposite direction. The coloured areas represent the Fermi-Dirac distribution function of each reservoir. To each process that is represented exists a process in the opposite direction that is however exponentially suppressed since it decreases the entropy. 
Tunneling events in each system differ by a charging energy $E_C$ when the other quantum dot either is empty or occupied. Thus, sequences involving the empty state and the simultaneous occupation of the two dots lead to the transfer of the energy $E_C$ and to directed electron motion in the conductor.}
\end{figure}
\begin{figure}
\includegraphics[width=\linewidth,clip]{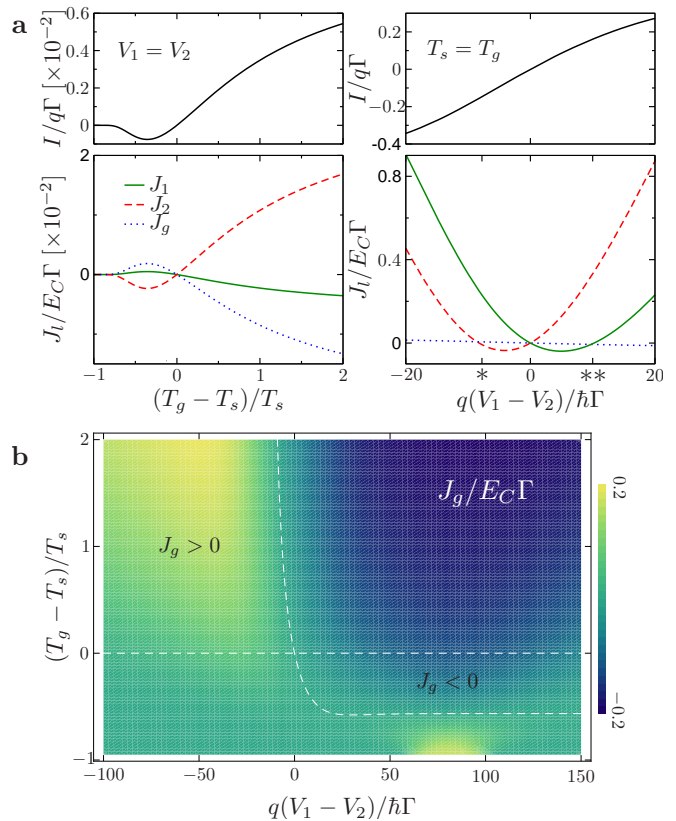}
\caption{\label{currs} Charge and heat currents as functions of temperature and voltage differences. (a) The direction of the currents can be tuned by changing the sign of the temperature gradient (left panel) or the applied bias voltage (right panel). At the points marked with $\ast$ and $\ast\ast$, the conditions $E_{s0}=qV_2$ and $E_{s0}=qV_1$ are satisfied. Between them, heat flows in opposite directions in the two terminals of the conductor, so one of its reservoirs is cooled down. Beyond them, Joule heat becomes dominant in the conductor. (b) Driving the conductor far enough from equilibrium, heat will flow from the coldest to the hottest system. In the delimited regions, heat flows into the gate ($J_g>0$) being $T_g>T_s$, and viceversa. Parameters: $\Gamma_{\!ln}=\Gamma$, except $\Gamma_{\!11}=0.1\Gamma$, $kT_1=kT_2=5\hbar\Gamma$, $q^2/C_i=20\hbar\Gamma$, $q^2/C=50\hbar\Gamma$, $\varepsilon_{\rm u}=\varepsilon_{\rm d}=0$, $V_g=V_1$.}
\end{figure}

If the two terminals in the transport system are at the same voltage, $V_1=V_2$, and temperature, $T_1 = T_2 = T_s$, only sequences that correlate the tunneling of an electron between different leads in the conductor with a charge fluctuation in the gate quantum dot (completing cycles ${\cal C}_\pm$ as shown in Fig.~\ref{bas}) contribute to break detailed balance. 
In the process, the transferred electron gains (${+}$) or loses (${-}$) an energy $E_C$ from or into the gate, respectively.
The entropy produced in the system by such cycles is 
\be
\Delta S_\pm=\pm E_C\left(\frac{1}{T_s}-\frac{1}{T_g}\right).
\ee
Processes that reduce entropy are exponentially suppressed. If, for instance, $T_s<T_g$, only processes where the electron increases its energy when traversing the quantum dot will contribute effectively to the current, and vice versa, as sketched in Fig.~\ref{proc}. Quite intuitively, heat will flow from the hottest to the coldest system. 
The probability  to transfer a particle from the left to the right leads by absorbing an energy $E_C$ is proportional to the product of the involved tunneling rates, $\Gamma_{10}\Gamma_{21}$, while for the reversed process (from right to left) one finds $\Gamma_{11}\Gamma_{20}$. The relevant sequences are sketched in Figure \ref{proc}. When these two products are different, a stationary current will flow in the unbiased conductor in a direction determined by the asymmetry of the tunneling rates:
\be
\label{currv0}
I =q\frac{\Gamma_{11}\Gamma_{20}-\Gamma_{10}\Gamma_{21}}{(\Gamma_{10}+\Gamma_{20})(\Gamma_{11}+\Gamma_{21})}\frac{J_g}{E_C}. 
\ee
A detailed derivation of this result can be found in Appendix~\ref{sec:flucgen}.
The charge current $I$ is proportional to the heat flux $J_g$ through the gate. 
Here $q$ is the charge of the electron and $E_C$ plays the role of the energy quantum. The close relation between these two currents in the absence of a voltage bias can be appreciated in the left panel of Fig. \ref{currs}a. Interestingly, in \eqref{currv0} the properties of the gate system are only contained in the heat current, $J_g$.

In this situation (when the conductor is unbiased and the gate is at a different temperature)  energy conservation also requires that heat currents are conserved. In contrast, when applying a finite voltage, the sum of heat currents is non-vanishing due to dissipation of Joule heat, as seen in the right panel of Fig. \ref{currs}a. In the linear regime, the charge response to temperature differences is related to the heat currents driven by voltage through Onsager relations~\cite{butcher}. Then, the direction of the electronic motion in the conductor changes its sign with the reversal of the temperature gradient or, correspondingly, the extraction or injection of heat from the conductor is determined by the sign of the generated current. This behaviour has been discussed for refrigeration by Brownian motors\cite{vandenbroeck}. Remarkably, the non-equilibrium state induced by the applied voltage allows to find regions where the flow of heat between the two systems is reversed, so, counter intuitively, it flows from the coldest to the hottest system, as seen in Fig. \ref{currs}b. 

\section{Efficiency}
\label{sec:efficiency}
As discussed above, our system can be used to transform heat flowing from a hot environment into electric current at zero power. In order to be transformed into useful work, a load system is needed. In other words, the current has to flow against a finite potential, $\Delta V$. Then, we can define the efficiency of the heat to current conversion, $\eta$, as the ratio of the obtained power, $P=I\Delta V$, to the absorbed heat, $J_g$. In what follows, we will consider that the gate system is at a higher temperature than the conductor, $T_g>T_s$. 
In our case, the heat absorbed from the gate system can contribute to processes that carry electrons in both directions, thus reducing the current, or to processes where
electrons tunnel back and forth between the quantum dot and the same reservoir in the conductor. 
This kind of processes involve heat transfer into the conductor, but do not contribute to the charge current in the desired direction, thus reducing the efficiency. 
As a result, the efficiency is limited by the tunneling prefactor appearing in Eq. \eqref{currv0}.

The contribution of these undesired processes will be negligible in the limiting case where $\Gamma_{l0},\Gamma_{r1}\gg\Gamma_{r0},\Gamma_{l1}$, for different leads $l$, $r$ of the conductor. In such an energy selective configuration, 
an electron that tunnels into the conductor quantum dot from lead $l$ can only be transmitted to lead $r$ after absorbing an energy $E_C$ from the gate, or tunnel back to $l$ without exchanging energy with the gate. The latter process is spurious, i.e. it does not contribute to the charge nor the heat currents, so it does not affect the efficiency. Thus, every time that an energy $E_C$ is absorbed from the gate, a charge $q$ is transferred in a given direction. Expressed differently, we can say that the gate system will not lose heat until an electron has been transferred from one lead to the other in the conductor. If an electron is transferred in the opposite direction, an energy $E_C$ is returned to the gate. The two currents are then related only by their quanta, 
\be
\label{propcurr}
\frac{I}{q}=-\frac{J_g}{E_C}. 
\ee
The proportionality between charge and heat carried by the same particle flow has been discussed 
to imply high thermoelectric efficiency~\cite{esposito}. Remarkably, our device achieves this property for crossed currents: a charge current flowing along the conductor and a heat current flowing through the gate.

\begin{figure}
\includegraphics[width=\linewidth,clip]{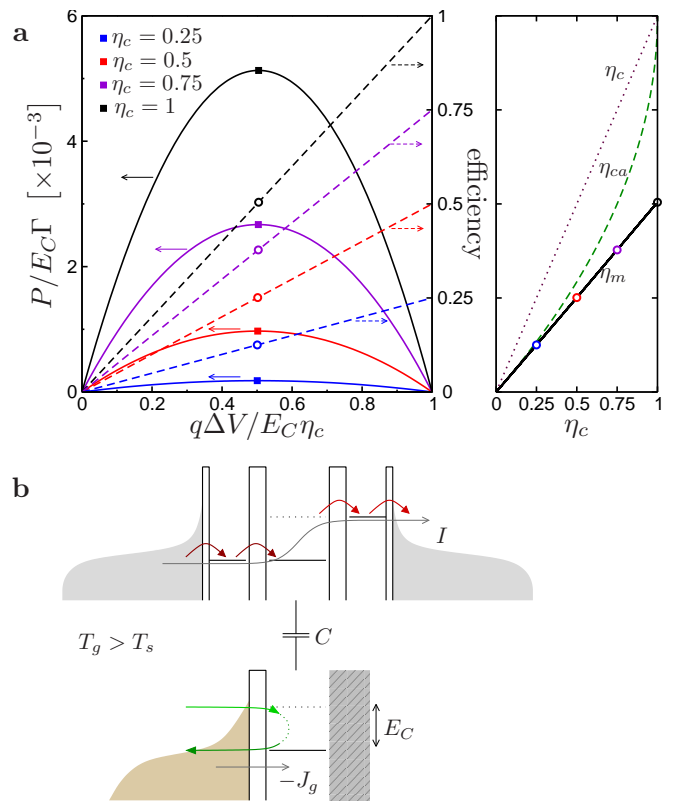}
\caption{\label{eff} Efficiency. (a)  
Power (solid lines) of the heat generated current and efficiency (dashed lines) of the heat to charge current conversion as a function of voltage for $kT_s=5\hbar\Gamma$ for different gate temperatures. The efficiency grows linearly up to the stopping potential, $V_0$, where Carnot efficiency is achieved. In the right panel, the efficiency at maximum power is plotted and compared with the Carnot and the Curzon-Ahlborn efficiencies.
Same parameters as in Fig.~\ref{currs}, except $\Gamma_{\!11}=\Gamma_{\!20}=0$. (b) Triple quantum dot system proposed to work as an optimal heat to charge current converter. The side quantum dots act as energy filters.}
\end{figure}

For our converter, the efficiency reduces to the simple expression 
\be
\eta(\Delta V)=\frac{q\Delta V}{E_C}.
\ee
Though it depends linearly on voltage, the Carnot efficiency, $\eta_c=1-T_s/T_g$, represents an upper limit which cannot be surpassed~\cite{callen}. By increasing the bias one always arrives at a point where the transfer of an electron does not produce entropy. At this point, the reversed processes (an electron tunnels from $r$ to $l$ by dissipating an energy $E_C$ into the gate) are equally probable and the current vanishes, as discussed in detail in Appendix~\ref{stc}. In our configuration, such stopping potential corresponds to $\Delta V=V_0=E_C\eta_c/q$, so Carnot efficiency is achieved, 
\be
\eta(V_0)=\eta_c, 
\ee
as shown in the left panel of Fig.~\ref{eff}a.

However, no power can be extracted from a heat engine working at Carnot efficiency. Note that, at $\Delta V=V_0$, no charge or heat current flows through the system, which is in equilibrium despite the applied voltage and temperature gradient~\cite{humphrey}. Hence it is more useful to discuss the efficiency at the point of maximum power extraction, $\eta_m$~\cite{curzonAhlborn,esposito}. As can be seen in Fig.~\ref{eff}a, it approaches the Curzon-Ahlborn efficiency, $\eta_{ca}=1-\sqrt{T_s/T_g}$, for small temperature differences so $\eta_m=\eta_c/2+O(\eta_c^2)$. As the temperature difference increases, $\eta_m$ becomes dependent on the configuration, in particular on the temperature of the conductor: the lower $T_s$, the closer is $\eta_m$ to the Carnot efficiency. In the range of validity of our simple model, $kT_s,kT_g\gg\hbar\Gamma$, where $k$ is the Boltzman constant, the efficiency at maximum power is maintained around $\eta_c/2$ far from the linear regime.

Though the energy selective configuration discussed above might be difficult to find in a single quantum dot system, it can be achieved in a triple quantum dot structure, where the outer quantum dots play the role of zero dimensional contacts~\cite{bryllert,corr}, 
as depicted in Fig.~\ref{eff}b. If, for instance, the energy level of the left (right) quantum dot is in resonance with the energy $E_{s0(1)}$, electrons can only be transferred from left to right by gaining an energy $E_C$ in the central quantum dot, or in the opposite direction by losing it. In order to avoid heat leakage from the interaction with the electrons in the outer quantum dots, it is required that the capacitance associated with their tunneling barriers is large, $C_{L}, C_R\gg C_1,C_2$. i.e. their charging energy is negligible.

\section{Discussion}
\label{sec:discussion}
We have identified a mechanism to generate directed electrical motion in a quantum dot system by the electrostatic coupling to a fluctuating gate at a different temperature. Energy exchange is quantized and depends only on the geometric capacitance of the quantum dots. 
In the optimal configuration, the ratio of charge to heat current is determined solely by the ratio of the charge to the energy quanta. Then, our device can be proposed as a solid state environmental energy to current converter of high efficiency.
Decoupling of the direction of energy flow and the direction of electron motion permits structures of multiple pairs of dots transferring heat in parallel increasing the total available power.

We introduce a mechanism based on well known phenomena as the quantization of energy levels and the physics of single charge Coulomb interaction~\cite{qdots} in a simple system which is experimentally available~\cite{bridge,hubel}. Temperature differences can be generated on mesoscopic scales~\cite{molenkamp,molenkamp2,molenkamp3} and thus our proposal is within experimental reach even with
present day structures. Exporting these ideas to other mesoscopic systems opens new possibilities for highly efficient solid state thermoelectric devices.

\acknowledgments
We would like to thank Ivar Martin for a stimulating discussion. This work was supported by MaNEP, the Swiss NSF and the strep project NANOPOWER (FP7/2007-2013) under grant agreement no. 256959. 


\appendix

\section{Electrostatic energies}
\label{sec:elener}

Considering the capacitance associated with every tunneling junction, the charge of each quantum dot, $Q_\alpha$, is given by
\be
Q_\alpha=\sum_{i_\alpha} C_{i_\alpha}(\phi_\alpha-V_{i_\alpha})+C(\phi_\alpha-\phi_\beta),
\ee
where $\phi_{\alpha}$ and $\phi_\beta$ are the electrostatic potential in each quantum dot, and $V_{i_\alpha}$ is the voltage of the reservoirs $i_\alpha$ to which the quantum dot is coupled, as sketched in Fig.~\ref{bas}. We denote by the indices $s$ and $g$ the quantum dot coupled to reservoirs 1 and 2, and the one coupled to the gate reservoir, respectively. We obtain the electrostatic energy in the quantum dot system for a given charge distribution:
\be
U(Q_s,Q_g)=\sum_\alpha\int_0^{Q_\alpha}dQ_{\alpha}'\phi_\alpha.
\ee
The relevant quantity is the change of energy in the quantum dot system when an electron tunnels through a tunnel junction thus modifying its charge. This is the charging energy. For the processes involving the empty system, $U_{s0}=U(1,0)-U(0,0)$ and $U_{g0}=U(0,1)-U(0,0)$, they read:
\bea
U_{s0}=\frac{q}{C\tilde C}\left(\frac{q}{2}C_{\Sigma g}+C_{\Sigma g}\sum_{i=1}^2C_iV_i+CC_gV_g\right)\\
U_{g0}=\frac{q}{C\tilde C}\left(\frac{q}{2}C_{\Sigma s}+C_{\Sigma s}C_gV_g+C\sum_{i=1}^2C_iV_i\right).
\eea
Additional energy is required when the other dot is already occupied, so $U_{s1}=U(1,1)-U(0,1)=U_{s0}+E_C$ and $U_{g1}=U(1,1)-U(0,1)=U_{g0}+E_C$, with $E_C=2q^2/\tilde C$. 
We recall here the definitions $C_{\Sigma s}=C_1+C_2+C$, $C_{\Sigma g}=C_g+C$, and $\tilde C=(C_{\Sigma s}C_{\Sigma g}-C^2)/C$.
Note that all the energies $U_{\alpha n}$ depend on the voltage of the three reservoirs. Thus, the effect of each system acting as a gate on the other one is included. $E_C$ determines the quantized energy which can be tranferred from the gate dot to the conductor. If the capacitive coupling of the two systems is sufficiently strong, $E_C$ can be of the order of the charging energy of the uncoupled quantum dots.

\section{Determining the quantum of energy}
\label{sec:ec}

\begin{figure}[t]
\begin{center}
\includegraphics[width=0.6\linewidth]{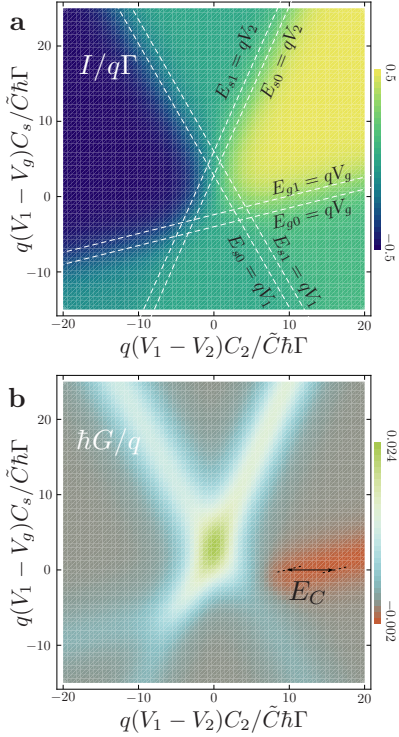}
\end{center}
\caption{\label{currcond} \fontfamily{phv} (a) Charge current when applying finite voltages to the conductor and to the gate. Dashed lines are plotted denoting the coincidence of the energies $E_{\alpha n}$ with the Fermi energies of the leads. The areas enclosed by them have a well defined charge occupation. (b) Differencial conductance, $G=\partial I/\partial (V_1-V_2)$. In the region $V_1>V_2$ and $E_{g0}<qV_g<E_{g1}$, the occupation of the gate quantum dot induces negative differential conductance. The width of such a plateau is determined by the quantum of transferred energy, $E_C$. Here, $C_s=C_1+C_2$. Parameters: $\Gamma_{\!ln}=\Gamma$, except $\Gamma_{\!11}=0.1\Gamma$, $kT_s=kT_g=5\hbar\Gamma$, $q^2/C_i=20\hbar\Gamma$, $q^2/C=50\hbar\Gamma$, $\varepsilon_s=\varepsilon_g=0$, $V_g=V_1$.}
\end{figure}

When the position of the discrete levels crosses the Fermi energy of the reservoirs to which they are coupled, the average charge occupation of the quantum dot system changes. In the regions in between two of these steps, the current presents a series of plateaus, as can be seen in Fig.~\ref{currcond}a. This is known as Coulomb blockade. In this regime, transport spectroscopy shows regions where the charge occupation is well defined resembling the Coulomb diamonds as a function of gate voltage and source-drain bias~\cite{qdots}. In our case, the interaction with the gate is mediated by a quantum dot, whose occupation modifies the stability diagram: Coulomb interaction between charges in each quantum dot takes the level out of the conduction window, thus avoiding charge transport when $E_{g0}<qV_g$.

These features are more visible in the differential conductance, $G=\partial I/\partial (V_1-V_2)$, as shown in Fig.~\ref{currcond}b. In the region where $V_1>V_2$ and $E_{g0}<qV_g<E_{g1}$, the gate quantum dot becomes occupied, thus increasing the energy that electrons need to be transferred through the conductor. Therefore, charge current through the conductor is reduced leading to negative differential conductance. The width of this region is determined by the quantum of transferred energy, $E_C$.

\section{Fluctuation generated transport}
\label{sec:flucgen}

We are interested in the ability to generate a finite current between two reservoirs that are at the same voltage and temperature in the presence of fluctuations in a side coupled sytem whose temperature is different. In our conductor system, this correspond to imposing the conditions $V_1=V_2=V_s$ and $T_1=T_2=T_s$ to the conductor. Then, both reservoirs share the same distribution function. For simplicity, let us define $f_{\alpha n}^-=f((E_{\alpha n}-qV_\alpha)/kT_\alpha)$ and $f_{\alpha n}^+=1-f_{\alpha n}^-$, where the index $\alpha=\{s,g\}$ refers to each system. We can express the rates as $\Gamma_{\alpha n}^\pm=\Gamma_{\alpha n}f_{\alpha n}^\pm$, with $\Gamma_{sn}=\Gamma_{1n}+\Gamma_{2n}$. 
If we consider the terms in Eq. \eqref{eqi2} separately, we get
\bea
&&\!\!\!\!\!\!\!\!\Gamma_{20}^+\bar\rho_{10}-\Gamma_{20}^-\bar\rho_{00}\nonumber\\[-2.5mm]
&&\\[-2.5mm]
&&=\frac{\Gamma_{20}\Gamma_{s1}\Gamma_{g0}\Gamma_{g1}}{\gamma^3}\left(f_{g0}^-f_{s1}^-f_{s0}^+f_{g1}^+-f_{s0}^-f_{g1}^-f_{s1}^+f_{g0}^+\right)\nonumber\\
&&\!\!\!\!\!\!\!\!\Gamma_{21}^+\bar\rho_{11}-\Gamma_{21}^-\bar\rho_{01}\nonumber\\[-2.5mm]
&&\\[-2.5mm]
&&=\frac{\Gamma_{21}\Gamma_{s0}\Gamma_{g0}\Gamma_{g1}}{\gamma^3}\left(f_{s0}^-f_{g1}^-f_{s1}^+f_{g0}^+-f_{g0}^-f_{s1}^-f_{s0}^+f_{g1}^+\right).\nonumber
\eea
Substracting these two expressions and introducing the explicit form of the Fermi-Dirac distribution, we find the charge current generated in the conductor by the fluctuations in the gate:
\bea
I&=&q\frac{(\Gamma_{10}\Gamma_{21}-\Gamma_{11}\Gamma_{20})\Gamma_{g0}\Gamma_{g1}}{8\gamma^3}\nonumber\\[-2.5mm]
&&\\[-2.5mm]
&&\!\!\!\!\!\!\times\sinh\left[\frac{E_C}{2}\left(\frac{1}{kT_g}-\frac{1}{kT_s}\right)\right]\prod_{\alpha,n}\cosh^{-1}\frac{E_{\alpha n}-qV_\alpha}{2kT_\alpha},\nonumber
\eea
for $\alpha=\{s,g\}$ and $n=\{0,1\}$.
In the same way, from Eq. \eqref{jg}, we can write the expression for the heat flow between the two systems:
\bea
J_g&=&-E_C\frac{\Gamma_{s0}\Gamma_{s1}\Gamma_{g0}\Gamma_{g1}}{8\gamma^3}\nonumber\\[-2.5mm]
&&\\[-2.5mm]
&&\!\!\!\!\!\!\times\sinh\left[\frac{E_C}{2}\left(\frac{1}{kT_g}-\frac{1}{kT_s}\right)\right]\prod_{\alpha,n}\cosh^{-1}\frac{E_{\alpha n}-qV_\alpha}{2kT_\alpha}.\nonumber
\eea
We can see that the two currents are proportional to each other, satisfying Eq.~(\ref{currv0}).
Then, if for instance $T_g>T_s$, heat will flow from the gate to the conductor. The energy dependent tunneling asymmetry, $\Gamma_{10}\Gamma_{21}-\Gamma_{11}\Gamma_{20}$, will determine the preferred direction for the electron motion that will allow a net flow of electrons in the conductor.

\section{Selective tunneling configuration}
\label{stc}
If we consider the special case where $\Gamma_{11}=\Gamma_{20}=0$, for any voltage and temperature configuration, the expression for the heat current \eqref{jg} can be further simplified:
\be
\label{jgsel}
J_g=-E_C\frac{\Gamma_{10}\Gamma_{21}\Gamma_{g0}\Gamma_{g1}}{\gamma^3}\left(f_{10}^-f_{g1}^-f_{21}^+f_{g0}^+-f_{g0}^-f_{21}^-f_{10}^+f_{g1}^+\right).
\ee
In the same way, we find:
\bea
J_1&=&\frac{E_{s0}-qV_1}{E_C}J_g\\
J_2&=&-\frac{E_{s1}-qV_2}{E_C}J_g
\eea
for the heat flowing through each terminal in the conductor. Considering that $\Gamma_{s0}^i=\Gamma_{10}^i$ and $\Gamma_{s1}^i=\Gamma_{21}^i$ in Eqs.  \eqref{stat} and \eqref{eqi2}, we can easily rewrite the charge current:
\be
\label{i2sel}
I=q\frac{\Gamma_{10}\Gamma_{21}\Gamma_{g0}\Gamma_{g1}}{\gamma^3}\left(f_{10}^-f_{g1}^-f_{21}^+f_{g0}^+-f_{g0}^-f_{21}^-f_{10}^+f_{g1}^+\right).
\ee
By comparing Eqs. \eqref{jgsel} and \eqref{i2sel}, we verify that the heat and charge currents are proportional for any applied voltage, with a constant of proportionality determined by the ratio of the energy and charge quanta, as expressed in \eqref{propcurr}.
Note that the total heat current corresponds to the Joule heat: $J_1+J_2+J_g=(V_1-V_2)I$.

Making use of the property of the Fermi functions $1- f(x)=e^xf(x)$, we can rewrite the Fermi factor between brackets in \eqref{jgsel} and \eqref{i2sel} as 
\be
\label{fst}
f_{10}^-f_{g1}^-f_{21}^+f_{g0}^+\left(1-e^{\frac{E_{s0}-qV_1}{kT_1}-\frac{E_{s0}-qV_2}{kT_2}}e^{E_C\left(\frac{1}{kT_g}-\frac{1}{kT_2}\right)}\right).
\ee
From \eqref{fst}, it is straightforward to see that, when the affinities of the conductor fulfill the condition 
\be
\frac{E_{s0}-qV_1}{kT_1}-\frac{E_{s0}-qV_2}{kT_2}=E_C\left(\frac{1}{kT_2}-\frac{1}{kT_g}\right),
\ee
all the heat and charge currents will vanish, so the system is in equilibrium in spite of the voltage and temperature differences. In the case when the two terminals in the conductor are at the same temperature, $T_1=T_2=T_s$, we find that the stopping potential is
\be
\label{stopV}
qV_0=q(V_2-V_1)=E_C\left(1-\frac{T_s}{T_g}\right).
\ee
In the energy converter configuration, when $T_g>T_s$, the factor in the right hand side of \eqref{stopV} coincides with the Carnot efficiency of the system, $\eta_c$.
In this particular case, the applied voltage cancels the fluctuation generated charge current while Joule heat cancels the heat flow between the two systems.

\end{document}